\documentclass[aps,prl,twocolumn]{revtex4}
\usepackage{amssymb}

\usepackage{graphicx}

\begin{document}

\title{Spin-density wave Fermi surface reconstruction in underdoped YBa$_2$Cu$_3$O$_{6+x}$}
\author{N.~Harrison}
\affiliation{National High Magnetic 
Field Laboratory, Los Alamos National Laboratory, MS E536,
Los Alamos, New Mexico 87545}
\date{\today}

\begin{abstract}
We consider the reconstruction expected for the Fermi surface of underdoped YBa$_2$Cu$_3$O$_{6+x}$ in the case of a collinear spin-density wave with a characteristic vector ${\bf Q}=(\pi[1\pm2\delta],\pi)$, assuming an incommensurability $\delta\approx$~0.06 similar to that found in recent neutron scattering experiments. A Fermi surface possibly consistent with the multiple observed quantum oscillation frequencies is obtained. From the low band masses expected using this model as compared with experiment, a uniform enhancement of the quasiparticle effective mass over the Fermi surface by a factor of $\approx$~7 is indicated. Further predictions of the Fermi surface topology are made, which may potentially be tested by experiment to indicate the relevance of this model to underdoped YBa$_2$Cu$_3$O$_{6+x}$.
\end{abstract}

\maketitle

The recent discovery of magnetic quantum oscillations in high $T_{\rm c}$ superconductors provides a unique opportunity to access the Fermi surface topology and the properties of the quasiparticles undergoing pair formation~\cite{doiron1,yelland1,bangura1,jaudet1,sebastian1,vignolle1,singleton1,audouard1,sebastian2}. The reports of multiple small pockets of carriers~\cite{sebastian1,singleton1,audouard1} and a negative Hall coefficient~\cite{bangura1,leboeuf1} in the underdoped systems YBa$_2$Cu$_3$O$_{6+x}$ and YBa$_2$Cu$_4$O$_8$  suggest a possible reconstruction of the Fermi surface by a density-wave order parameter.  In this paper, we calculate the Fermi surface topology considering the case of a collinear spin-density wave with a modulation vector ${\bf Q}=(\pi,\pi[1\pm2\delta])$, where $\delta\approx$~0.06 is provided by recent elastic~\cite{haug1} and inelastic~\cite{stock1} neutron scattering experiments on underdoped~YBa$_2$Cu$_3$O$_{6+x}$. In this model, we use the values of $\delta$ and ratio $V_{\rm s}/t\ll$~1 (of the spin modulation potential to the in-plane hopping parameters) which represent the case of a conventional spin-density wave, rather than stripes of the form proposed in Refs.~~\cite{tranquada1,kivelson1}.

\begin{figure}[htbp!]
\centering
%\vspace{-8mm}
\includegraphics[width=0.45\textwidth]{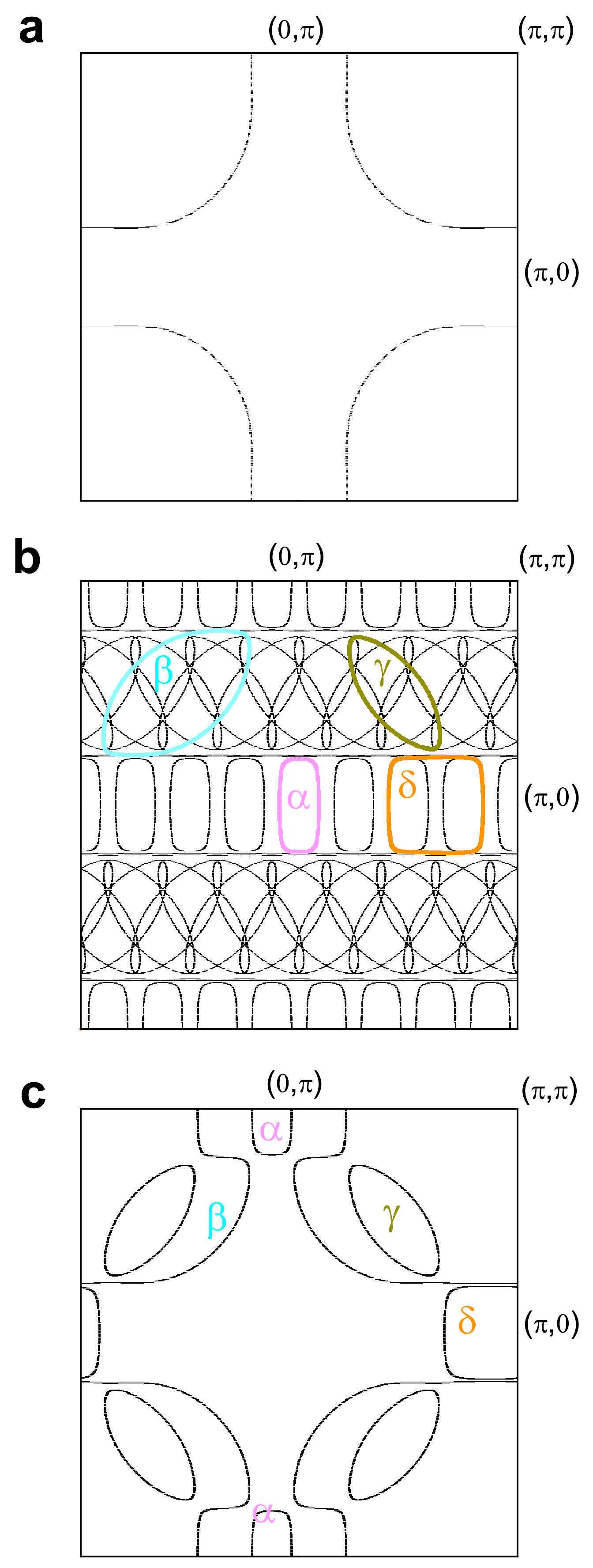}
%\vspace{-7mm}
\caption{{\bf a} Unreconstructed Fermi surface of YBa$_2$Cu$_3$O$_{6+x}$ for $p=$~0.1 according to Eqn.~(\ref{tightbinding}). {\bf b} Reconstructed Fermi surface, as described in the text. The four prominent orbits are depicted in color (online). {\bf c} Notional Fermi surface corresponding to the eigenvalues of a reduced  3~$\times$~3 matrix containing only the terms in the top-left-hand-corner of Eqn.(\ref{matrix}), which produces the same orbits as in ({\bf b}).}
\label{Fermisurface}
%\vspace{-3mm}
\end{figure}

We begin by considering the unreconsructed tight binding approximation
\begin{eqnarray}\label{tightbinding}
\varepsilon_k=\varepsilon_0+2t_{10}[\cos(ak_x)+\cos(bk_y)]+2t_{11}[\cos(ak_x\nonumber\\+bk_y)+\cos(ak_x-bk_y)]+2t_{20}[\cos(2ak_x)\nonumber\\+\cos(2bk_y)]+2t_c\cos ck_z
\end{eqnarray}
originally proposed for YBa$_2$Cu$_3$O$_{6+x}$ by Andersen {\it et al.}~\cite{andersen1}. In this approximation, we choose $t_{10}=-$~380~meV, $t_{11}/t_{10}=-$~0.32 and $t_{20}/t_{10}=$~0.16~\cite{andersen1,millis1}, while the number of holes is tuned by $\varepsilon_0$. We introduce an interlayer hopping term ($2t_c\cos ck_z$) to model the deep corrugation of the otherwise cylindrical Fermi surface section reported in recent studies~\cite{audouard1,sebastian2}. We further neglect the ortho-II potential~\cite{elfimov1}, following reports of quantum oscillations in non-ortho-II ordered samples~\cite{singleton1}, and the effect of bilayer splitting~\cite{andersen1,note0}. Angle-resolved photoemission spectroscopy experiments on surface K-deposition-treated ortho-II ordered YBa$_2$Cu$_3$O$_{6.5}$ yield a Fermi surface whose location in $k$-space is consistent with these assumptions for Eqn.~(\ref{tightbinding}) for hole dopings near $p=$~0.1~\cite{hossain1}. We plot the corresponding Fermi surface for $k_z=\pm\pi/2c$ in Fig.~\ref{Fermisurface}a.

For guidance on the appropriate choice of $\delta$, we turn to recent neutron scattering experiments on underdoped YBa$_2$Cu$_3$O$_{6+x}$. Magnetic field-dependent elastic scattering experiments~\cite{haug1} find evidence for long range antiferromagnetism with $\delta\approx$~0.055 in samples of similar composition ($x=$~0.45) to those of the lowest composition ($x=$~0.49) in which quantum oscillations are observed~\cite{sebastian2}. Meanwhile, inelastic scattering experiments performed on samples of nominally the same composition ($x\approx$~0.5) as those in which multiple Fermi surface pockets are reported~\cite{sebastian1,audouard1}, find incommensurate magnetic excitations at energies as low as 6~meV at $\delta\approx$~0.06~\cite{stock1} in the absence of an applied field. 

We therefore consider the possibility that spin-density wave ordering occurs near $\delta\approx$~0.06 in YBa$_2$Cu$_3$O$_{6.5}$ in sufficiently strong magnetic fields. If it is collinear, the Fermi surface reconstruction is determined by multiple translations of $\varepsilon_{\rm k}$ by $\pm n{\bf Q}$ (each of which we denote $\varepsilon_{{\bf k}+n{\bf Q}}$). As in Cr~\cite{fawcett1,lomer1}, this leads to a hierarchy of gaps of order $2\Delta_m\sim2V^m_{\rm s}/t^{m-1}$ opening at the crossing points of bands $\varepsilon_{{\bf k}+n{\bf Q}}$ and $\varepsilon_{{\bf k}+{n\pm m}{\bf Q}}$, where $V_{\rm s}$ is the amplitude of the spin potential and $t\sim t_{10}$. 
Provided $V_{\rm s}\ll t$, these gaps become vanishingly small for large $m$, implying that there is little difference between the orbits obtained in an incommensurate model with an irrational $\delta$ and those obtained using a commensurate model with a rational $\delta$ of similar value. We therefore choose to consider a rational $\delta$, since it enables a complete picture of the reconstructed Fermi surface to be obtained from the eigenvalues of a matrix~\cite{millis1}. Given the limited precision of inelastic neutron scattering experiments on YBa$_2$Cu$_3$O$_{6.5}$~\cite{stock1}, we make the convenient choice of $\delta=\frac{1}{16}=$~0.0625, and compute the eigenvalues of the 16~$\times$~16 matrix
\begin{equation}\label{matrix}
H=\left( \begin{array}{cccccc}
\varepsilon_{\bf k} & V_{\rm s} & V_{\rm c} & \dots & V_{\rm c} & V_{\rm s}\\
V_{\rm s} & \varepsilon_{{\bf k}+{\bf Q}} & V_{\rm s} & \dots  & 0 & V_{\rm c}\\
V_{\rm c} & V_{\rm s} & \varepsilon_{{\bf k}+2{\bf Q}} & \dots & 0 & 0\\
\vdots & \vdots & \vdots & \ddots & \vdots & \vdots \\
V_{\rm c} & 0 & 0 & \dots & \varepsilon_{{\bf k}+14{\bf Q}} & V_{\rm s}\\
V_{\rm s} & V_{\rm c} & 0 & \dots & V_{\rm s} & \varepsilon_{{\bf k}+15{\bf Q}} \end{array} \right).
\end{equation}
Additional terms ($V_{\rm c}$) are introduced in Eqn.~(\ref{matrix}) to accommodate a possible charge modulation at $2{\bf Q}$~\cite{millis1}, which often occurs in spin-density wave systems~\cite{fawcett1}. 

Figure~\ref{Fermisurface}b shows the in-plane Fermi surface cross-section at $k_z=\pm\pi/2c$ corresponding to the 16 eigenvalues of Eqn.~(\ref{matrix}) for the case of a simple collinear spin-density wave in which $V_{\rm c}=0$. Here, $V_{\rm s}$ is adjusted to a value of 0.15~$\times~t_{10}$ to yield an electron pocket ($\alpha$) of similar $k$-space area to that reported experimentally~\cite{leboeuf1}, while $\varepsilon_0$ is adjusted to conserve the hole filling at $p=$~0.1. The full paramagnetic Brillouin zone shown is 16 times larger than the antiferromagnetic Brillouin zone. Four primary orbits ($\alpha$, $\beta$, $\gamma$ and $\delta$) are expected for this reconstructed Fermi surface, and are indicated in color in Fig.~\ref{Fermisurface}b. The smallness of the higher order gaps $2\Delta_{m>1}$ implies that these four orbits  can be approximately reproduced by considering the eigenvalues of a reduced 3~$\times$~3 matrix containing only on the terms in the top-right-right-most corner of Eqn.~(\ref{matrix}). These are shown in Fig.~\ref{Fermisurface}c.
The multitude of very small hole orbits in Fig.~\ref{Fermisurface}b (resulting from the overlap of many bands with relative translations of $m{\bf Q}$ where $m>2$) are too small to have filled Landau levels in magnetic fields of the magnitude relevant for quantum oscillations~\cite{doiron1, jaudet1,sebastian1,vignolle1,singleton1,audouard1,sebastian2}. Furthermore, the gaps between them ($2\Delta_m\sim2V^m_{\rm s}/t^{m-1}\ll2V_{\rm s}$) are small enough to be completely broken through (having magnetic breakdown tunneling probabilities of $\approx$~1~\cite{shoenberg1}).

\begin{figure}[htbp!]
\centering
%\vspace{-8mm}
\includegraphics[width=0.45\textwidth]{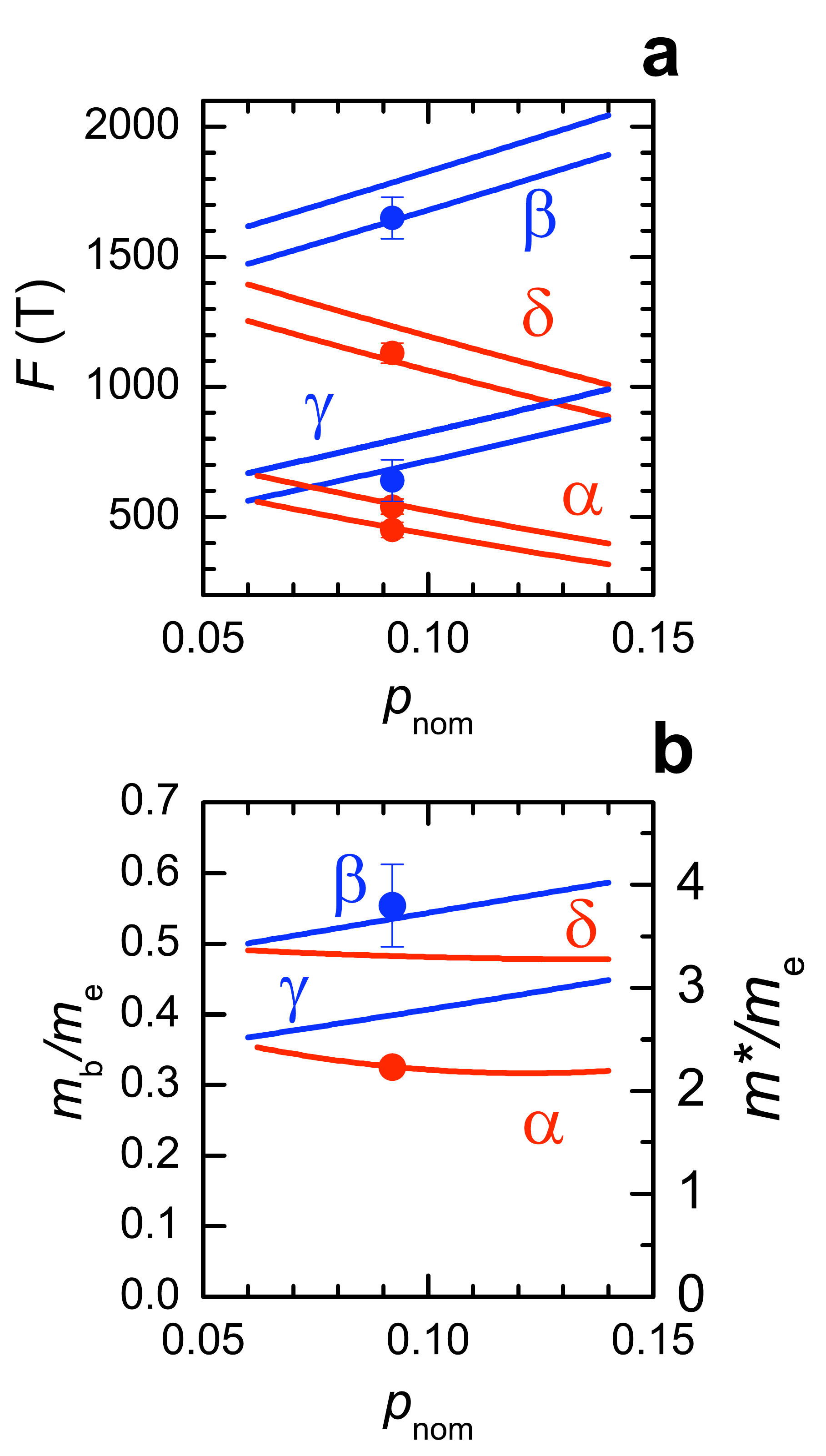}
%\vspace{-7mm}
\caption{{\bf a} A comparison of calculated quantum oscillation frequencies (lines) plotted for a range of $p$ with those observed experimentally (circles, with error bars at 2$\sigma$ level). A finite value of $t_{\rm c}\approx$~8~meV yields two parallel lines (i.e. `belly' and `neck' frequencies) for each Fermi surface section. Red and blue curves and symbols correspond electrons and holes respectively. Best agreement is obtained for $p=$~0.092. Experimental values for all the frequencies are taken from Ref.~\cite{audouard1}, where the wide magnetic field range enables accurate estimates, with the exception of $F_\beta$ taken from Ref.~\cite{sebastian1}. {\bf b} Comparison of the calculated band masses $m_{\rm b}$ (lines and left-hand-axis) with those $m^\ast$ reported experimentally (circles and right-hand-axis). A renormalization of $t_{10}$ from 390~meV to $\approx$~57~meV is required in order to align the calculated and experimental values, suggesting a possible uniform mass enhancement of $m^\ast/m_{\rm b}\approx$~7 across the Fermi surface. Effective mass estimates are taken from Refs.~\cite{sebastian1,sebastian2}.}
%\vspace{-3mm}
\label{Fandm}
\end{figure}

The corresponding quantum oscillation frequencies $F_i=(A_i/A_{\rm BZ})h/eab$ for each of the four primary orbits are plotted as a function of $p$ in Fig.~\ref{Fandm}a (where $\delta$ and $V_{\rm s}$ are fixed) to allow for possible discrepances in hole doping estimates~\cite{liang1,doiron1}. The frequencies are calculated from their $k$-space areas $A_i$ using the Onsager relation~\cite{shoenberg1}, where $A_{\rm BZ}$ is the area of the paramagnetic Brillouin zone and $a$ and $b$ are the in-plane lattice constants.  A value of $t_c\approx$~8~meV is required to account for the deep corrugation of (the otherwise cylindrical) Fermi surface detected in recent experiments~\cite{audouard1,sebastian2,note1}, yielding two (maximum `belly' and minimum `neck') frequencies for each Fermi surface section.

An appealling feature of the present Fermi surface model is that the smallest detectable quantum oscillation frequency predicted is the electron pocket (referred to here as the $\alpha$-pocket following the nomenclature introduced in Ref.~\cite{sebastian1}), similar to experimental findings~\cite{doiron1,jaudet1,sebastian1,audouard1,sebastian2}.  Furthermore, this pocket has the lightest effective mass (see Fig.~\ref{Fandm}b) and is the only one (of the four) not requiring magnetic breakdown for its observation, suggesting that it will dominate both the quantum oscillation spectrum in amplitude~\cite{sebastian1,audouard1} and the Hall coefficient in intermediate magnetic fields~\cite{leboeuf1}. 
The remaining three prominent orbits ($\beta$, $\gamma$ and $\delta$ in Fig.~\ref{Fermisurface}b) require varying degrees of magnetic breakdown to be observed, for which we make a rough estimate of their magnetic breakdown probabilities using
\begin{equation}\label{probability}
p_i\approx\exp\bigg(-\frac{4\Delta^2_iB}{\hbar^2\omega_{{\rm c},i}^2F_i}\bigg).
\end{equation}
The subscript $i$ refers to a particular orbit of frequency $F_i$ and cyclotron frequency $\omega_{{\rm c},i}=eB/m_{{\rm b},i}$, requiring tunneling through a gap $2\Delta_i$ to be observed~\cite{shoenberg1,note2}. Of the three orbits ($\beta$, $\gamma$ and $\delta$), the hole pocket denoted $\gamma$ requires magnetic breakdown through only a third order gap of order $2\Delta_3\sim2V^3_{\rm s}/t^2_{10}$ (plus less important higher order gaps), giving rise to a tunneling probability of order $p_\gamma\sim$~98~\% for $B\approx\mu_0H=$~50~T. The frequency of $\approx$~640~T reported by Audouard {\it et al}~\cite{audouard1,note3} provides a possible candidate for this orbit (see Fig.~\ref{Fandm}a). 
Since the $\delta$ and $\beta$ orbits require magnetic breakdown tunneling through the larger first order gap $2\Delta_1\sim2V_{\rm s}$, their quantum oscillation amplitudes are expected to be significantly weaker than those originating from the $\alpha$ and $\gamma$ orbits. The estimated magnetic breakdown probabilities are $p_\delta\sim$~2~\% and $p_\beta\sim$~8~\% respectively. Possibly consistent with these greatly reduced tunneling probabilities, weak features of similar frequency ($\approx$~1130~T and $\approx$~1650~T in Fig.~\ref{Fandm}a) to those in the model are reported in Refs.~\cite{sebastian1,audouard1}, although the harmonics of $F_\alpha$ are expected to occur nearby in frequency~\cite{audouard1}.

We find that the ratio of $m^\ast_\beta$ to $m^\ast_\alpha$ in Fig.~\ref{Fandm}b found experimentally is similar to that of $m_{{\rm b},\beta}$ to $m_{{\rm b},\alpha}$ predicted by the model, suggesting a possible uniform enhancement of the quasiparticle effective mass over the Fermi surface  of $\approx$~7 (relative to the bandstructure estimate~\cite{andersen1}). On considering this renormalization factor over the entire Fermi surface, the total summed effective mass of all Fermi surface sections within the antiferromagnetic Brillouin zone becomes $m^\ast_{\rm AFM}\approx$~7~$m_{\rm e}$ (where $m_{\rm e}$ is the free electron mass), compared to $m^\ast_{\rm P}\approx$~11~$m_{\rm e}$ for the single large orbit of the unreconstructed paramagnetic Fermi surface. Using these estimates, we obtain electronic coefficients of the heat capacity of $\gamma_{\rm AFM}=$~1.46~$\times(m^\ast_{\rm AFM}/m_{\rm e})=$~10~mJmol$^{-1}$K$^{-2}$~\cite{sebastian1} and $\gamma_{\rm P}=$~1.46~$\times(m^\ast_{\rm P}/m_{\rm e})=$~16~mJmol$^{-1}$K$^{-2}$ for the reconstructed and unreconstructed Fermi surfaces respectively.

\begin{figure}[htbp!]
\centering
%\vspace{-8mm}
\includegraphics[width=0.45\textwidth]{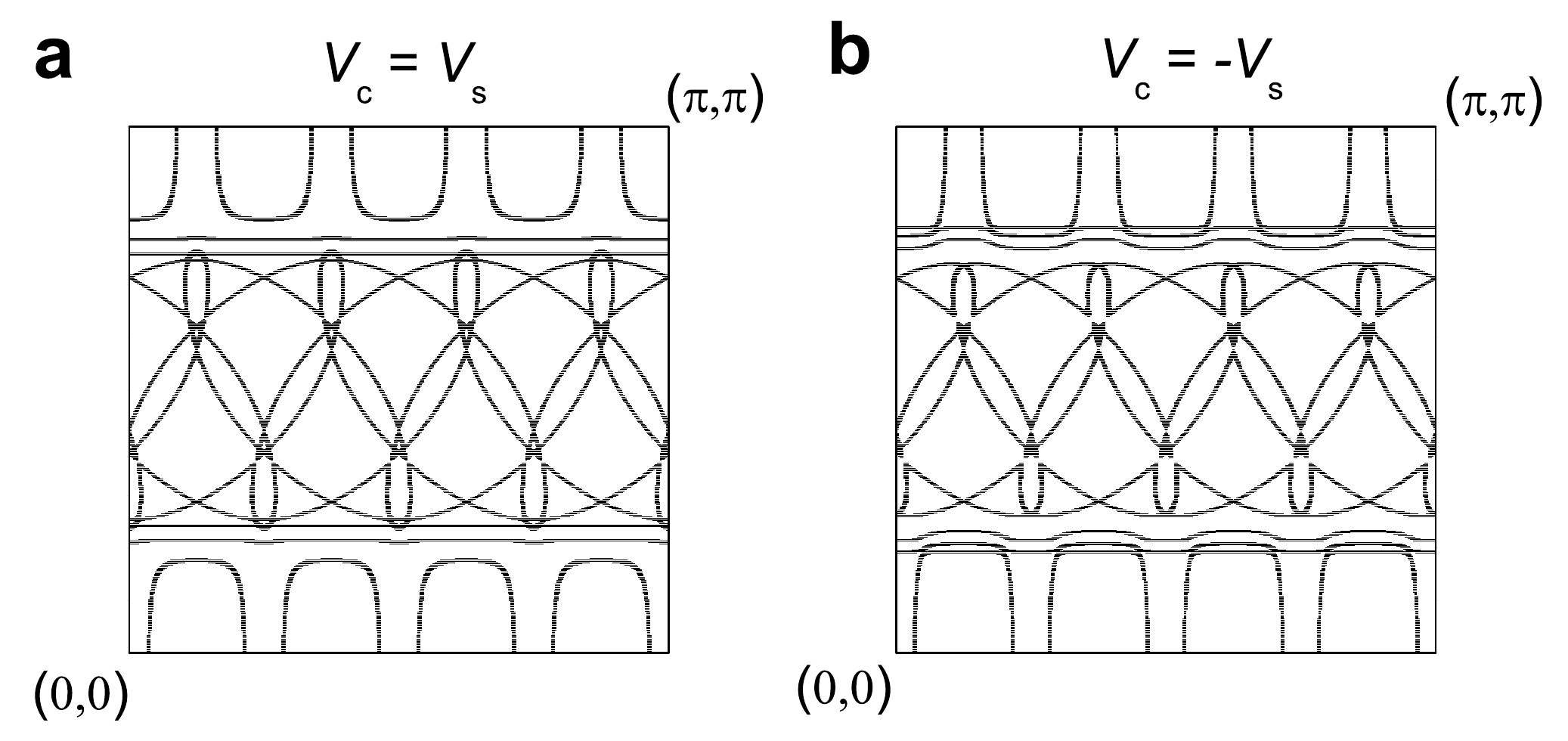}
%\vspace{-7mm}
\caption{The Fermi surfaces (shown for $\frac{1}{4}$ of the paramagnetic Brillouin zone) obtained on including a significant charge modulation potential at $2{\bf Q}$, with $V_{\rm c}=V_{\rm s}$ in ({\bf a}) and $V_{\rm c}=-V_{\rm s}$ in ({\bf b}).}
%\vspace{-3mm}
\label{charge}
\end{figure}

One notable caveat in assessing the applicability of the spin-density wave model to YBa$_2$Cu$_3$O$_{6+x}$ presents itself in Fig.~\ref{Fandm}a on comparing the experimental frequencies with those predicted in the model. Best agreement is obtained for $p=$~0.092~\cite{liang1}, which falls short of that $p\approx$~0.098~$\pm$~0.001 estimated from the lattice parameter c~\cite{doiron1,liang1}. While some uncertainty in $t_{20}/t_{10}$ or $t_{11}/t_{10}$ in Eqn.~(\ref{tightbinding}) or our neglect of the ortho-II potential (which occurs at $8{\bf Q}$ here) can likely account for some of the discrepancy, our neglect of the $2{\bf Q}$ charge modulation potential may be more significant. On including $V_{\rm c}$ in the simulations in Fig.~\ref{charge}, the electron pocket shrinks or expands depending on whether $V_{\rm c}$ is positive or negative, suggesting that improved consistency with the nominal hole doping level could be achieved by the simultaneous adjustment of $V_{\rm s}$ and $V_{\rm c}$.

In summary, by using a value $\delta\approx$~0.06 for the incommensurability parameter found in recent neutron scattering experiments, we find that the existence of a collinear spin density wave in suitably strong magnetic fields would yield a reconstructed Fermi surface consistent with the observed multiple carrier pockets in underdoped YBa$_2$Cu$_3$O$_{6+x}$~\cite{sebastian1,audouard1}. Four major orbits are predicted, with magnetic breakdown being an important factor for all but the smallest electron pocket.

A key feature of this model is that the anticipated shape of the electron pocket is a rounded rectangle with an aspect ratio $>\approx$~2 and its long axis parallel to $k_y$ (orthogonal to the incommensurate spin modulation direction)$-$ something that can be tested by dual-axis angle-dependent magnetic quantum oscillation or magnetoresistance experiments. A small electron pocket that is found to be circular in cross-section (or have an approximate fourfold symmetry) in experiments would be unreconcilable with the present collinear density wave model and would instead have different implications. Such a pocket would then suggest an evolution of $\delta$ rather abruptly to a value significantly greater than 0.06 in strong magnetic fields~\cite{note4}, or a helical or spiral spin-density wave~\cite{note5}, or an entirely different form of Fermi surface topology in YBa$_2$Cu$_3$O$_{6+x}$ (possibly unrelated to the magnetic diffraction peaks). 
If experiments detect a Fermi surface topology similar to that yielded by the collinear spin density wave model considered here, indications would be that such a form of order could be chiefly responsible for Fermi surface reconstructuion in strong magnetic fields in underdoped YBa$_2$Cu$_3$O$_{6+x}$. Given the small value of $V_{\rm s}$ that we consider in this model, a further implication would be that the observed quasiparticle mass enhancement has an origin mostly extrinsic to this form of collinear
density wave.

This work is supported by the US Department of Energy, the National Science Foundation and the State of Florida. The author acknowledges helpful comments from S.~E.~Sebastian.


\begin{thebibliography}{99}

\bibitem{doiron1} 
N.~Doiron-Leyraud {\it et al.}, Nature {\bf 447}, 565 (2007).

\bibitem{yelland1} 
E.A.~Yelland {\it et al.}, Phys. Rev. Lett. {\bf 100}, 047003 (2008).

\bibitem{bangura1}
A.~F.~Bangura {\it et al.}, Phys. Rev. Lett. {\bf 100}, 046004 (2008).

\bibitem{jaudet1} C.~Jaudet {\it et al.}, preprint arXiv:0711.3559 (2008).

\bibitem{sebastian1}
S.~E.~Sebastian {\it et al.}, Nature {\bf 454}, 200 (2008).

\bibitem{vignolle1}
B.~Vignolle {\it et al.}, Nature {\bf 455}, 952 (2008).

\bibitem{singleton1} J.~Singleton, R.~D.~McDonald, and S.~Cox, Physica B {\bf 404}, 350 (2009).

\bibitem{audouard1} A.~Audouard {\it et al.}, preprint arXiv:0812.0458 (2008).

\bibitem{sebastian2} S.~E.~Sebastian (preprint, 2009).

\bibitem{leboeuf1}
D.~LeBoeuf {\it et al.}, Nature {\bf 450}, 533 (2007).

\bibitem{haug1} D.~Haug {\it et al.}, Magnetic field enhanced incommensurate magnetism in the underdoped high-temperature superconductor YBa$_2$Cu$_3$O$_{6.45}$ (preprint, 2008).

\bibitem{stock1} C.~Stock {\it et al.}, Phys, Rev. B {\bf 69}, 014502 (2004); C.~Stock {\it et al.}, Phys. Rev. B {\bf 71}, 024522 (2005).

\bibitem{kivelson1} S.~A.~Kivelson {\it et al.}, Rev. Mod. Phys. {\bf 75}, 1201 (2003).

\bibitem{tranquada1} J.~M.~Tranquada {\it et al.}, Nature {\bf 375}, 561 (1995).

\bibitem{andersen1} O.~K.~Andersen {\it et al.}, Phys. Chem. Solids {\bf 56}, 1573 (1995).

\bibitem{millis1}
A.~J.~Millis and M.~R.~Norman, Phys. Rev. B {\bf 76}, 220503 (2007).

\bibitem{elfimov1} I.~S.~Elfimov, G.~A.~Sawatsky, and A.~Damascelli, Phys. Rev. B {\bf 77}, 060504 (2008).

\bibitem{note0} We consider the simplest case of a spin-density wave that couples bonding and antibonding states, leading to two sets of degenerate reconstructed Fermi surfaces in which the effects of bilayer splitting are averaged out.

\bibitem{hossain1} M.~A.~Hossain {\it et al.}, Nature Phys. {\bf 4}, 527 (2008).

\bibitem{lomer1} W.~M.~Lomer, Proceedings of the
International Conference on Magnetism (Nottingham, UK, 1964).

\bibitem{fawcett1}
E.~Fawcett, Rev. Mod. Phys. {\bf 60}, 209 (1988).

\bibitem{shoenberg1} D.~Shoenberg, {\it Magnetic oscillations in metals} (Cambridge University Press,
Cambridge 1984).

\bibitem{liang1} R.~Liang, D.~A.~Bonn, and W.~N~Hardy, Phys. Rev. B {\bf 73}, 180505 (2006).

\bibitem{note1} Because the experimental effective masses are renormalized by a factor of $\approx$~7 compared to the model (in which $t_{10}=-$~390~meV), the experimental estimates of $t_{\rm c}^\ast\approx$~1~meV in Refs.~\cite{audouard1,sebastian2} must be considered renormalized down by a similar factor.

\bibitem{dimov1} I.~Dimov {\it et al.}, Phys. Rev. B {\bf 78}, 134529 (2008).

\bibitem{note2} For convenience, we have equated the effective Fermi energy in the standard expression for magnetic breakdown~\cite{shoenberg1} with $\varepsilon_{{\rm F},i}=\hbar eF_i/m_{{\rm b},i}$ for a particular orbit.

\bibitem{note3} In Ref.~\cite{audouard1}, this orbit has a different magnetic field-dependent attenuation compared to the $\approx$~450~T and $\approx$~540~T frequencies attributed to the $\alpha$ pocket~\cite{sebastian2}, suggestive of a different effective mass or scattering rate, which would normally indicate a different Fermi surface section.

\bibitem{note4} In the present collinear model, an incremental increase in either $V_{\rm s}$ or $\delta$ worsens the level of agreement between the area of the predicted electron pocket and the $\alpha$ frequency reported in experiments. Not until $\delta>\approx$~0.1 is it possible once again to obtain an electron pocket with an area compatible with experiments using a collinear model~\cite{millis1}.

\bibitem{note5} For a helical spin-density wave, the Fermi surface can be obtained by solving a 2~$\times$~2 matrix~\cite{sebastian1}, which reduces the number of crossing points of the translated bands, yielding an electron pocket in the shape of a rounded square.












\end{thebibliography}
\end{document}